\documentstyle[aps,amssymb,twocolumn]{revtex}

\begin{document}
\author{David D. Reid}
\address{Department of Physics and Astronomy\\
Eastern Michigan University, Ypsilanti, MI 48197}
\title{Introduction to causal sets: an alternate view of spacetime structure }
\maketitle

\begin{abstract}
{\bf Abstract}. This paper provides a thorough introduction to the causal
set hypothesis aimed at students, and other interested persons, with some
knowledge of general relativity and nonrelativistic quantum mechanics. I
elucidate the arguments for why the causal set structure might be the
appropriate structure for a theory of quantum gravity. The logical and
formal development of a causal set theory as well as a few illuminating
examples are also provided.
\end{abstract}

\section{Introduction}

When studying general relativity, students often find that two of the most
compelling topics, cosmology and black holes, lead directly to the need for
a theory of quantum gravity. However, not much is said about quantum gravity
at this level. Those who search for more information will find that most
discussions center around the two best know approaches: canonical
quantization [1] and superstring theories [2]. This paper seeks to introduce
the problem of quantum gravity in the context of a third view, causal sets,
which has emerged as an important concept in the pursuit of quantum gravity.

The causal set idea is an hypothesis for the structure of spacetime. This
structure is expected to become apparent for extremely tiny lengths and
extremely short times. This hypothesis, in its current form, has grown out
of an attempt to find an appropriate structure for a physical theory of
quantum gravity. There is a long tradition of the importance of causality in
relativity. Many of the issues faced when confronting the problem of quantum
gravity bring considerations of time and causality to the forefront.

There are many approaches to quantum gravity. Usually, these approaches go
through cycles of rapid progress, during which times an approach will appear
very promising, followed by (sometimes long) periods of slow, or no,
progress. The causal set approach has gone through these cycles as well,
although to a lesser extent than some, with early work by Finkelstein [3],
Myrhiem [4], 't Hooft [5], and Sorkin [6]. The recent upswing of interest in
causal sets was ignited by a paper written in the late 1980s [7]. Since the
hope is that causal sets will lead to a working model for quantum gravity,
it seems appropriate to begin by describing the problem of quantum gravity
in general. The basic ideas behind the causal set approach and some of the
progress that has been forged in recent years will be discussed in sections
III - VI.

\section{The problem of quantum gravity}

\subsection{What is quantum gravity?}

The question of what one means by ``quantum gravity'' is not a simple
question to answer for the obvious reason that we do not yet have a complete
understanding of quantum gravity. Hence, the answers to this question are
both short and long and perhaps as numerous as the number of approaches
attempting to solve the problem. Most physicists agree that by ``gravity''
we mean Einstein's theory of general relativity (and possibly a few modified
versions of it). General relativity most popularly interprets gravitation as
a result of the geometrical structure of spacetime. The geometrical
interpretation fits because the theory is formally cast in terms of metrical
structure $g_{\mu \nu }$ on a manifold $M$.

There is somewhat less agreement on the meaning of ``quantum.'' At first
glance, it seems odd that there would be less agreement on the aspect with
which we have much more experience. On the other hand, however, the fact
that we have only been able to perform weak-field experimental tests of
general relativity leaves us with much less information to debate. Our
experience with quantum mechanics tells us that the deviations from
classical physics it describes are important when dealing with size scales
on the order of magnitude of an atom and smaller. Is there a natural size
scale at which we expect the predictions of general relativity to be
inaccurate requiring a new more fundamental theory?

The scale at which theories become important is set by the values of the
fundamental parameters related to the processes being described. For
example, the speed of light $c$ is the fundamental constant that determines
the velocity scale for which relativistic effects (special relativity) are
appreciable. Likewise, Planck's constant $\hslash $, among others, sets the
scale for systems that must be described by quantum mechanics. The
fundamental constants that are relevant to a theory of quantum gravity are
the speed of light, Planck's constant, and the universal gravitation
constant $G$. These three quantities combine to form the length and time
scales at which classical general relativity break down: 
\begin{equation}
\left. 
\begin{array}{l}
\ell _P=\left( G\hslash /c^3\right) ^{1/2}\sim 10^{-35}\text{ m} \\ 
t_P=\ell _P/c\sim 10^{-44}\text{ s,}
\end{array}
\right.  \label{1}
\end{equation}
where $\ell _P$ is called the Planck length and $t_P$ is called the Planck
time.

Size, however, is only one part of what makes a theory ``quantum.''
Consider, once again, the atom. If we dig deeper than just size and ask why
quantum effects are important for atoms, the answer is that a relatively
small number of states are occupied (or excited). This fact is more commonly
stated in reverse as a correspondence principle requiring that quantum
mechanics merge with classical mechanics in the limit of large quantum
numbers, that is, a large number of occupied states. It is this latter point
that truly characterizes quantum behavior. A quantum theory must therefore
enumerate and describe the states of a system in such a way that the known
classical behavior emerges for large numbers of states.

What, then, do we mean by ``quantum gravity?'' In this paper, my working
definition is that

\begin{quotation}
{\it quantum gravity is a theory that describes the structure of spacetime
and the effects of spacetime structure down to sub-Planckian scales for
systems containing any number of occupied states.}
\end{quotation}

\noindent In the above definition, the ``effects of spacetime structure''
include not only the phenomenon of gravitational attraction, but also any
implications that the spacetime dynamics will have for other interactions
that take place within this structure.

\subsection{Why do we need quantum gravity?}

\subsubsection{The Einstein field equations}

The content of the Einstein field equations of general relativity, 
\begin{equation}
G_{\mu \nu }=\kappa T_{\mu \nu },  \label{2}
\end{equation}
suggests the need for a quantum mechanical interpretation of gravity [8].
Here $G_{\mu \nu }$ is the Einstein curvature tensor representing the
curvature of space-time, $T_{\mu \nu }$ is the energy-momentum tensor
representing the source of gravitation, while $\kappa $ is just a coupling
constant between the two. The energy-momentum content of spacetime is
already known to be a quantum operator from other fundamental theories such
as quantum electrodynamics (QED). We have confidence in the reliability of
this interpretation because, despite the fact that QED may have flaws
(discussed below), it has led to extremely accurate agreement between theory
and experiment [9]. Since energy-momentum is a quantum operator whose
macroscopic version is intimately related to macroscopic spacetime
structure, it seems a good working hypothesis that its quantum mechanical
version should correspond to a structure of space and time appropriate in
the quantum mechanical regime.

\subsubsection{Black holes}

A {\it black hole} is the final stage in the evolution of massive stars.
Black holes are formed when the nuclear energy source at the core of a star
is exhausted. Once the nuclear fuel has run out, the star collapses. If the
remaining mass of the star is sufficiently high, no known force can halt the
collapse. General relativity predicts that the stellar mass will collapse to
a state of zero extent and infinite density -- a {\it singularity}. In this
singular state there is no spatial extent, time has no meaning, and the
ability to extract any physical information is lost. This prediction may be
a message which tells us that a quantum theory of gravity is needed if we
are to truly understand the inner workings of black holes.

While it may be obvious that processes deep within black holes must be
treated in the framework of quantum gravity, it is less obvious that
processes well away from the singularity not only require quantum gravity,
but may also provide important clues to the form a theory of quantum gravity
should take. In 1975 Hawking [10] showed that black holes radiate thermally
with a blackbody spectrum. This finding, together with a previous conjecture
that the area of a black hole's event horizon can be interpreted as its
entropy [11], has shown that the laws of black hole mechanics are identical
to the laws of thermodynamics. This equivalence only comes about if we
accept the {\it identification} of the area of the black hole (actually $1/4$
of it) as its entropy. In traditional thermodynamics the concept of entropy
is best understood in terms of discrete quantum states; not surprisingly,
attempts to better understand the reasons for this area identification using
classical gravity{\it \ }fail. It is widely expected that only a quantum
mechanical approach will produce a satisfactory explanation [12]. For this
reason, black hole entropy is an important topic for most approaches to
quantum gravity [13].

\subsubsection{The early universe}

One of the many triumphs of relativistic cosmology is the explanation of the
observed redshift of distant galaxies as an expansion of the universe.
However, the universal expansion extrapolates backward to an early universe
that is infinitesimally small and infinitely dense -- the big bang
singularity. Here then, is another situation in which general relativity
predicts something it is not equipped to describe. It is fully expected that
events near the singularity were dominated by quantum mechanical influences
both {\it of} and {\it on} spacetime which necessarily affects the
subsequent evolution of the universe. Presently, cosmological implications
of the early universe are studied with the techniques of quantum cosmology
[14] which is the quantum mechanics of classical cosmological models. It has
been pointed out that quantum cosmology cannot be trusted except in very
specialized cases [15]. While we have good theories for doing quantum
mechanics on background spacetimes, the early universe problem requires a
theory for the quantum mechanics of spacetime itself -- a theory of quantum
gravity.

\subsubsection{Unification}

Throughout the history of physics, great strides have been made through the
unification of seemingly different aspects of nature. One of the most
prominent examples is Maxwell's unification of the laws of electricity and
magnetism. Einstein's theory of special relativity amounts to a unification
of Maxwell's electromagnetism and Newton's mechanics showing Newton's laws
to be merely a ``low speed'' approximation to a more accurate relativistic
dynamics. Following relativity theory, quantum mechanics was born. Soon
thereafter, Dirac unified quantum mechanics and special relativity giving
rise to modern quantum field theory.

With the above successful unifications behind us, we are left with the
present situation of having several fundamental forces known as the strong,
weak, and electromagnetic interactions as well as gravitation. Given the
benefits that we have reaped from past unifications it seems natural that
the search for deeper insight through unification should continue. The
recent success of the electroweak{\it \ }theory has confirmed the value of
this search. There are now some seemingly consistent models for the
unification of the strong and electroweak theories. The very fact that these
interactions can be mathematically unified in a manner consistent with
macroscopic observations suggests that a truly physical unification exists.

Gravity is the only fundamental force yet to have a consistent quantum
mechanical theory. It is widely believed that until such a quantum
mechanical description of gravity is attained, placing our understanding of
gravity on the same level as that of the other interactions, true
unification of gravity with the other forces will not be possible [8].

\subsection{The incompatibility between general relativity and quantum
mechanics}

For all of the interactions except gravity, our present theoretical
understanding of physics is such that systems interact and evolve within a
background spacetime structure. This background structure serves to tell us
how to measure distances and times. In general relativity it is the
spacetime structure itself that we must determine. This spacetime structure
then, acts both as the background structure for gravitational interactions
and as the dynamical phenomenon giving rise to this interaction. In general
relativity the structure of spacetime is determined by the Einstein field
equations (2). These field equations are, however, purely classical in that
they do not meet the requirements of a quantum theory as discussed in
section II.A above. The breakdown of general relativity near the singularity
of a black hole, or more accurately, the prediction of a singularity inside
of a black hole, is just one of many examples. Given that general relativity
was formulated prior to quantum mechanics, the fact that it does not meet
quantum mechanical requirements is not surprising.

The dual role of the metric tensor makes formulating a theory of quantum
gravity very different from the formulations of the other interactions. In
quantum gravity we must determine the spacetime structure that acts as
background to the classical structure of space and time that we have used to
understand all other phenomena. Furthermore, this ultimate background to
classical spacetime structure must also be dynamic because it is this
dynamics that will describe quantum gravity just as the dynamics of
classical spacetime describes general relativity. This latter point is the
key reason for the incompatibility between general relativity and quantum
mechanics. All of our successful experience is with quantum dynamics on a
spacetime structure, but we have had very little success handling the
quantum dynamics of spacetime structure.

This incompatibility challenges some of the most fundamental concepts in
physics. In field theory, we take as the source of the field some
distribution $T_{\mu \nu }$ (of charge, matter, energy, etc.). However, the
concept of a spatial distribution of charge, for example, has no meaning
apart from the knowledge of the spatial structure. Without the rules for how
to measure relative positions we cannot define the spatial distribution of
anything. In quantum mechanics, Schr\"{o}dinger's equation describes the
time evolution of the wave function $\Psi ({\bf r},t)$. The concept of time
evolution, however, requires an existing knowledge of temporal flow which
comes from the spacetime structure. As a final example, consider the concept
of interaction. We generally think of interactions in a manner intimately
connected with causality: an interaction precedes and causes an effect.
Causality, however, is a concept that can only be defined once the structure
of spacetime is known.

\section{The causal set hypothesis}

The above discussion implies the need for a spacetime structure that will
underpin the classical spacetime structure of general relativity. The causal
set hypothesis proposes such a structure. Causal sets are based on two
primary concepts: the discreteness of spacetime and the importance of the
causal structure. Below I discuss these two founding concepts in more detail.

\subsection{Spacetime is discrete}

The causal set hypothesis assumes that the structure of spacetime is
discrete rather than the continuous structure that physics currently
employs. Discrete means that lengths in three-dimensional space are built up
out of a finite number of elementary lengths $\ell _e$ which represents the
smallest allowable length in nature and the flow of time occurs in a series
of individual ``ticks'' of duration $t_e$ which represents the shortest
allowable time interval.

The idea that something which appears continuous is actually discrete is
very common in physics and everyday life. Any bulk piece of matter is made
up of individual atoms so tightly packed that the object appears continuous
to the naked eye. Likewise, any motion picture is constructed of a series of
snapshots so rapidly paced that the movie appears to flow continuously.

Why hold a similar view of spacetime? There are many arguments for a
discrete structure. The most familiar ones are related to electrodynamics.
Here, I will first try to motivate the idea of discreteness by considering
the electromagnetic spectrum. The electromagnetic spectrum gives us the
range for the frequencies and wavelengths of electromagnetic radiation, or,
photons. Of the many interesting aspects of this spectrum, here let's focus
on that fact that it is a continuous spectrum of infinite extent. Current
theory predicts that the allowed frequencies of photons extend continuously
from zero to infinity. The relation $E=h\nu $ implies that a photon of
arbitrarily large frequency has arbitrarily large energy. The local
conservation of energy suggests that such infinite energy photons should not
exist. If one adopts the (somewhat controversial) view that what cannot
physically exist should not be predicted by theory, there ought to be a
natural cutoff of the electromagnetic spectrum corresponding to a maximum
allowed frequency. Since the frequency of a photon is the inverse of its
period, a discrete temporal structure provides a natural cutoff in that the
minimum time interval $t_e$ implies a maximum frequency $\nu _{\max }=1/t_e$.

Because of relativity any argument for discrete time is also an argument for
discrete space. Nevertheless, a similar argument for discrete space can be
given in terms of wavelength. The electromagnetic spectrum, being
continuous, allows arbitrarily small wavelengths. The deBroglie relation $%
p=h/\lambda $ implies that a wavelength arbitrarily close to zero
corresponds to a photon of arbitrarily large momentum. The local
conservation of momentum suggests that photons of infinite momentum should
not exist. The minimum length implied by a discrete spacetime structure
provides a natural cutoff for the wavelength $\lambda _{\min }=\ell _e$.

In terms of QED, this problem can be seen in the fact that the infinite
perturbation series requires the existence of all the photons in the
electromagnetic spectrum. In this sense QED predicts the existence of these
photons of infinite energy-momentum. However, it is widely believed that the
perturbation series diverges. This divergence is generally overlooked
because QED is only a partial theory and not a complete theory of elementary
interactions (see Ch. 1 of Ref. 1). Therefore, we use QED under the
assumption that it is accurate for the phenomenon it was created to describe
and that some aspect of a more fundamental theory will eventually solve its
divergence problem. The causal set idea proposes that the aspect of more
fundamental theory that will naturally solve this divergence problem in QED
is a discrete structure of spacetime.

\subsection{Causal structure contains geometric information}

Spacetime consists of events $x^{\mu {}}=(x^0,x^1,x^2,x^3)=(ct,x,y,z)$, that
is, points in space at various times. At some events physical processes take
place. Processes that occur at one event can only be influenced by those
occurring at another event if it is possible for a photon to reach the
latter event from the earlier one. To capture the essence of what one means
by ``causal structure,'' consider the example of the flat Minkowski space of
special relativity. In flat spacetime, two events are said to be causally
connected and their spacetime separation $ds^2=g_{\mu \nu }dx^{\mu
{}}dx^{\nu {}}$ is called timelike if it is positive (using a $+---$
signature) and null if it is zero. Two events are not causally connected if
it is not possible for a photon from one event to arrive at the other; these
events cannot influence each other and in such cases $ds^2$ is negative and
referred to as spacelike. When we speak of the causal structure of a
spacetime, we mean the knowledge of which events are causally connected to
which other events. For a more general discussion of causal structure see
reference [16].

It is now well established that the causal structure of spacetime alone
determines almost all of the information needed to specify the metric [17,
18] and therefore the gravitational field tensor. The causal structure
determines the metric up to an overall multiplicative function called a
conformal factor. We say that two metrics $g_{\mu \nu }$ and $\widetilde{g}%
_{\mu \nu }$ are conformally equivalent if $\widetilde{g}_{\mu \nu }=\Omega
^2g_{\mu \nu }$, where $\Omega $ is a smooth positive function. Since all
conformally equivalent spacetimes have the same causal structure [19], the
causal structure itself nearly specifies the metric.

\subsection{Causal sets}

Lacking the conformal factor from the causal structure essentially means
that we lack the sense of scale which allows for quantitative measures of
lengths and volumes in spacetime. However, if spacetime is discrete, the
volume of a region can be determined by a procedure almost as simple as
counting the number of events within that region. Therefore, if nature
endows us with discrete spacetime and an arrow for time (the causal
structure), we have, in principle, enough information to build complete
spacetime metric tensors for general relativity. This combination of
discreteness and causal structure leads directly to the idea of a causal set
as the fundamental structure of spacetime.

A causal set may be defined as a set of events for which there is an order
relation $\prec $ obeying four properties:

\begin{enumerate}
\item  {\it transitivity}: if $x\prec y$ and $y\prec z$ then $x\prec z$;

\item  {\it non-circularity}: if $x\prec y$ and $y\prec x$ then $x=y$;

\item  {\it finitarity}: the number of events lying between any two fixed
events is finite;

\item  {\it reflexivity}: $x\prec x$ for any event in the causal set.
\end{enumerate}

The first two properties say that this ordered set is really a partially
ordered set, or {\it poset} for short. Specifically, non-circularity amounts
to the exclusion of closed timelike curves more commonly known as time
machines [20]. The finitarity of the set insures that the set is discrete.
The reflexivity requirement is present as a convenience to eliminate the
ambiguity of how an event relates to itself. In the present context of using
a poset to represent spacetime, reflexivity seems reasonable in that the
spacetime separation between an event and itself cannot be negative
requiring an event to be causally connected to itself. We can combine these
statements to give the following definition:

\begin{quote}
A{\it \ causal set is a locally finite, partially ordered set}.
\end{quote}

\section{The development of causal set theory}

With the conceptual foundation of causal sets clearly laid, let us now turn
to the issue of developing a physical and mathematical formalism which I
loosely refer to as {\it causal set theory}. The development of causal set
theory is still far from complete. In fact, it is even less developed than
some of the other approaches to quantum gravity such as superstring theory
and canonical quantization mentioned previously. For this reason, quantum
gravity by any approach is an excellent theatre for fine tuning ideas about
how theory construction should proceed from founding observations,
hypotheses, and assumptions. Hence, to better understand current thought on
how the development of causal set theory should proceed, I will first
discuss some of the ideas on theory development in general that have
influenced causal set research.

\subsection{Taketani stages}

In 1971 Mituo Taketani used Newtonian mechanics as a prototype to illustrate
his ideas on the development of physical theories [21]. According to
Taketani, physical theories are developed in three stages that he referred
to as the phenomenological, substantialistic, and essentialistic stages.

The phenomenological stage is where the initial observations occur that
place the existence and knowledge of the new phenomenon (or ``substance'')
on firm standing. In Ref. [21] this stage in the development of Newtonian
mechanics is associated with the work of Tycho Brahe who observed the
motions of the planets with unprecedented accuracy.

Within the substantialistic stage, rules that describe the new substance are
discovered; that is, we come to recognize ``substantial structure'' in the
new phenomenon. These rules would then play an important role in helping to
shape the final understanding. For Newtonian mechanics, Taketani associated
this stage with the work of Johannes Kepler who provided the well known
three laws of planetary motion.

In the essentialistic stage, ``the knowledge penetrates into the essence''
of the new phenomenon. This is the final stage when the full theory of this
new substance is known within appropriate limits of validity. Of course, the
work of Issac Newton himself represents this stage.

Even though Taketani used Newtonian mechanics, there are many examples to
which his ideas apply. Sakata used Taketani's philosophy to discuss the
development of quantum mechanics [22]. Similarly, the development of
electromagnetic theory falls neatly into Taketani's framework. The
phenomenological stage of electromagnetism could be associated with the work
of Benjamin Franklin and William Gilbert. The substantialistic stage is
nicely represented by the work of Michael Faraday and Hans Oerstead. The
essentialistic stage is then represented by James Maxwell's completion of
his famous equations.

Taketani realized that his three stages will not always apply identically to
the development of all physical ideas. Since we currently know of no
observed phenomena whose explanation clearly requires a complete theory of
quantum gravity, it is clear that the problem of quantum gravity is not
based on experimental observations. As a result of this fact, the
development of causal set theory largely skips the phenomenological stage.
Therefore, think of the development of causal sets in a two-step process
corresponding roughly to Taketani's substantialistic and essentialistic
stages. As a matter of terminology, note that the substantialistic stage, in
which phenomenology is described, plays the role of kinematics in Newtonian
mechanics, while in the essentialistic stage the full dynamics is developed.
Consequentially, I will refer to the two processes in the development of
causal set theory as ``kinematics'' and ``dynamics.''

\subsection{Causal set kinematics}

The kinematic stage concerns gaining familiarity with and further developing
the mathematics needed to describe causal sets. This mathematics primarily
falls under the combinatorial mathematics of partial orders [23]. These
techniques are not part of traditional physics training and have, therefore,
not been widely used to analyze physical problems. Moreover, research in
this branch of mathematics has been performed largely by pure
mathematicians; the problems they have chosen to tackle are generally not
those that are of most interest to physicists. What we need from the
kinematic stage are the mathematical techniques for how to extract the
geometrical information from the causal order (i.e., working out the
correspondence between order and geometry) and how to do the counting of
causal set elements that will allow us to determine spacetime volumes.

For an important, specific example of where causal set kinematics is needed,
consider the correspondence principle between spacetime as a causal set and
macroscopic spacetime. General relativity tells us that spacetime is a
four-dimensional Lorentzian manifold. If causal sets comprise the true
structure of spacetime they must produce a four-dimensional Lorentzian
manifold in macroscopic limits (such as a large number of causal set
elements). The mathematics of how we can see the manifold within the causal
set is a kinematic issue that must be addressed.

On the natural length scale of the causal set one does not expect to see
anything like a manifold. Trying to see a manifold on this scale is like
trying to read this article under a magnification that resolves the
individual dots of ink making up the letters. To discern the structure of
these dots we look upon them at a significantly different scale than the
size scale of the dots. Similarly, we need a mathematical change-of-scale in
order to extract the manifold structure from the causal set. This
change-of-scale is called {\it coarse-graining}.

Some insight into this issue can be gained by looking at the reverse problem
of forming a causal set from a given metric manifold $(M,g_{\mu \nu })$.
This is achieved by randomly sprinkling points into $M$. The order relation
of this set of points is then determined by the light-cone structure of $%
g_{\mu \nu }$. Since we need to ensure that every region of the spacetime is
appropriately sampled, that is, that highly curved regions are represented
equally well as nearly flat regions are, the sprinkling is carried out via a
Poisson process [24] such that the number of points $N$ sprinkled into any
region of volume $V$ is directly proportional to $V$. Using a
two-dimensional Minkowski spacetime, Fig. 1 provides a picture of such a
sprinkling at unit density $\rho =N/V=1$.

Since the causal sets generated by random sprinklings are only expected to
be coarse-grained versions of the fundamental causal set, their length and
time scales are not expected to be the fundamental length and time of
nature. Nevertheless, these studies of random sprinklings are important
because the founding ideas behind causal sets in Sec. III suggests that {\it %
a manifold }$(M,g_{\mu \nu })${\it \ emerges from the causal set }$C${\it \
if and only if an appropriately coarse-grained version of }$C${\it \ can be
produced by a unit density sprinkling of points into }$M$ [25]. This shows
us that an important problem in the development of causal set kinematics is
to determine how to appropriately form a coarse-graining of a causal set.

\subsection{Causal set dynamics}

The final stage in the development of causal set theory is the stage in
which we come to understand the full dynamics of causal sets. In this stage
we devise a formalism for how to obtain physical information from the
behavior of the causal set and how this behavior governs our sense of space
and time. Here we require something that might be considered a quantum
mechanical analog to the Einstein field equations (2). Since our present
framework for physical theories is based on a spacetime continuum, our
experience is of limited use to us in this effort. Despite this limitation,
one commonly used approach stands out as the best candidate for a dynamical
framework for causal sets. This method is most commonly known as the
path-integral formulation of quantum mechanics [26].

This path-integral technique seems best suited to causal sets because at its
core conception (a) it is a spacetime approach in that it deals directly
(rather than indirectly) with events; that is, we propagate a system from
one event configuration to another; and (b) it works on a discrete spacetime
structure. As currently practiced, the path-integral approach determines the
propagator $U(a^{\mu {}},b^{\nu {}})$ by taking all paths between the events 
$a^{\mu {}}$ and $b^{\nu {}}$ in a discretized time and summing over these
paths using an amplitude function $\exp (iS/\hslash )$: 
\begin{equation}
U(a^{\mu {}},b^{\nu {}})\sim \sum_{all\text{ }paths}\exp (iS/\hslash ),
\label{3}
\end{equation}
where $S$ is the action for a given path. Continuous spacetime enters in at
two places. In a continuum there are an infinite number of paths between two
events, ``all paths'' are generated by integrating over all intermediate
points between the two events; this is the ``integral'' part of
path-integration. Since each of these paths were discretized into a finite
number of points $N$, the second place where continuous spacetime is
recovered is to take the limit $N\rightarrow \infty $.

In a discrete setting the number of paths and the number of points along the
paths are truly finite. Hence, the final limit as well as the integration to
generate all paths are not performed. Since causal sets would not require
integration, calling this the ``path-integral formulation'' seems
inappropriate. This method essentially says that the properties of a system
in a given event configuration depend on a sum over all the possible paths
throughout the history of this system. Therefore, the alternative name for
this technique, the {\it sum-over-histories} approach, is better suited for
causal sets. The word ``histories'' is particularly appropriate because, as
mentioned earlier, we take the arrow of time to be fundamental.

There are several key issues that must be resolved before a
sum-over-histories formulation of causal sets can be completed. One such
issue involves the need to identify an amplitude function for causal sets
analogous to the role played by $\exp (iS/\hslash )$. Secondly, the required
formulation must do more than just propagate the system because the entire
dynamics must come from this formalism. The procedure outlined above is
presently inadequate for these purposes; a modified, or better, generalized
sum-over-histories method must be developed.

Perhaps the most significant advance along the dynamical front is the recent
development by Rideout and Sorkin of a general {\it classical} dynamics for
causal sets [27]. In this model, causal sets are grown sequentially, one
element at a time, under the governance of reasonable physical requirements
for causality and discrete general covariance. When a new element is
introduced, in going from an $n$-element causal set to an $(n+1)$-element
causal set, it is associated with a classical probability $q_n$ of being
unrelated to any existing element according to 
\[
\frac 1{q_n}=\sum\limits_{k=0}^n%
{n \choose k}%
t_k. 
\]
The primary restriction is that the $t_k\geq 0$; hence, there is a lot of
freedom with which different models can be explored. This framework has the
potential to teach us much about the needed mathematical formalism for
causal sets, the effects of certain physical conditions, and the classical
limit of the eventual quantum dynamics for causal sets.

\section{Illustrative examples}

To illustrate some of the points discussed above I present the 72 element
causal set shown in Fig. 2. The black dots represent the elements of the
causal set. The graphical form in which this causal set is shown is known as
a {\it spacetime-Hasse} diagram. The term ``Hasse diagram'' is borrowed from
the mathematical literature on posets [23]. Figure 2 is also a spacetime
diagram in the usual sense. The solid lines in the figure are {\it causal
links}, i.e., lines are only drawn between events that are causally related;
however, for clarity only those relations that are not implied by
transitivity (the links) are explicitly shown. The causal set shown has 15
time steps as enumerated along the right side of the figure. Hence, the
first time step at the bottom shows 7 ``simultaneous'' events.

\subsection{Volume}

As discussed in Sec. III, the causal set hypothesis is partially founded on
the fact that the causal structure of spacetime contains all of the
geometric information needed to specify the metric tensor up to a conformal
factor which prevents us from determining volumes. One example which shows
how, in principle, volumes might be extracted from the causal set has been
discussed by Bombelli [28]. Gerard 't Hooft has shown [5] that, in Minkowski
space, the volume $V$ of spacetime bounded by two causally connected events $%
a$ and $b$ is given by 
\begin{equation}
V=\frac{\pi \tau _{ab}^4}{24},  \label{4}
\end{equation}
where $\tau _{ab}$ is the proper time between events $a$ and $b$. We can
apply this expression to causal sets by relating $\tau _{ab}$ to the number
of links in the longest path between $a$ and $b$. The volume $V$ is then
identified with the number of elements in this region of spacetime.

Spacetime is dynamic, however. The above procedure of counting the number of
links between two events is subject to (perhaps large) statistical
fluctuations. Therefore, while it is believed that the expected proper time $%
<\tau >$ should be proportional to the number of links [29], the precise
relationship between them is yet unknown. Attempts to numerically determine
this relationship via computer simulations remain inconclusive [28].

\subsection{Coarse-graining}

While the labeling of Fig. 2 clearly suggests that it is a two-dimensional
example of a causal set, note that our physical sense of dimensionality
(given us by relativity) is intimately related to the manifold concept.
Although the causal set in Fig. 2 looks suspiciously regular it may not be
immediately obvious whether or not this set can be embedded into any
physically viable two-dimensional spacetime.

As an example of one form of coarse-graining, we can look at this causal set
on a time scale twice as long as its natural scale. Figure 3a is a
coarse-grained version of the causal set in Fig. 2 for which only even time
steps are shown and Fig. 3b is a subset of Fig. 2 showing events that only
occur at odd time steps. In both cases we find causal sets that clearly can
be embedded into two-dimensional Minkowski space. In a realistic situation
this fact would suggest that the fundamental causal set just might represent
a physically discrete spacetime.

\subsection{Dynamics}

As stated above, a sum-over-histories dynamical law for causal sets requires
the identification of an amplitude function. As an example, one could start
by considering an amplitude modeled after the familiar amplitude of the
continuum path-integral formulation in Eq. (3), i.e., $\exp (i\beta R)$.
Here, $\beta $ plays the role of $1/\hslash $ and $R$ plays the role of the
action $S$. In quantum field theory the oscillatory nature of this amplitude
causes problems that are sometimes bypassed by performing a continuation
from real to imaginary time (often referred to as a Wick rotation).
Similarly, it is convenient here to consider the case $\beta \rightarrow
i\beta $ giving an amplitude 
\begin{equation}
A=\exp (-\beta R),  \label{5}
\end{equation}
where we can take $R$, for example, to be the total number of links in the
causal set.

This model is interesting because the amplitude $A$, which acts as a weight
in the sum-over-causal sets, has the form of a Boltzmann factor $\exp
(-E/kT) $. The mathematical structure of the causal set dynamics then
becomes very similar to that of statistical mechanics. Studies of the
statistical mechanics of certain partially ordered sets have been performed
[30]. In the thermodynamic limit, these studies exhibit phase transitions
corresponding to successively increasing numbers of layers of the lattice
causing the poset to appear more and more continuous. In this analogy, the
thermodynamic limit corresponds to one macroscopic limit of causal set
theory in which the number of causal set elements goes to infinity. Such
results, therefore, are somewhat suggestive that an appropriate choice of
amplitude might indeed lead to the expected kind of continuum limit.

Another, more detailed, example of a quantum dynamics for causal sets that
exhibits the kind of interference effects that are absent from the classical
dynamics mentioned previously can be found in Ref. [31].

\section{Concluding Remarks}

In this paper we have tried to communicate the primary motivation and key
ideas behind the causal set hypothesis. Causal sets has emerged as an
important approach to quantum gravity having been found to impact other
approaches such as the spin network formalism [32].

Adding to the importance of the causal set approach is the fact that it led
Sorkin to predict a non-zero cosmological constant nearly a decade ago [25].
In light of recent findings in the astrophysics community [33, 34], this
result perhaps marks the only prediction to come out of quantum gravity
research that might be testable in the foreseeable future.

Before a final causal set theory can be constructed, much work remains.
Studies of random sprinklings in both flat and curved spacetimes, the
mathematics of partial orders, and the behavior of fields that sit on a
discrete substructure are just a few areas of needed investigation. Enough
progress on causal sets has been made, however, to establish the causal set
hypothesis as a very promising branch of quantum gravity research. Those
with further interest can find more detailed discussion of causal sets in
Refs. [35, 36].

\section{Acknowledgments}

The author would like to thank Drs. R. D. Sorkin, J. P. Krisch, N. L.
Sharma, and E. Behringer for helpful suggestions. Support of the Michigan
Space Grant Consortium is gratefully acknowledged.

\section{Figure Captions}

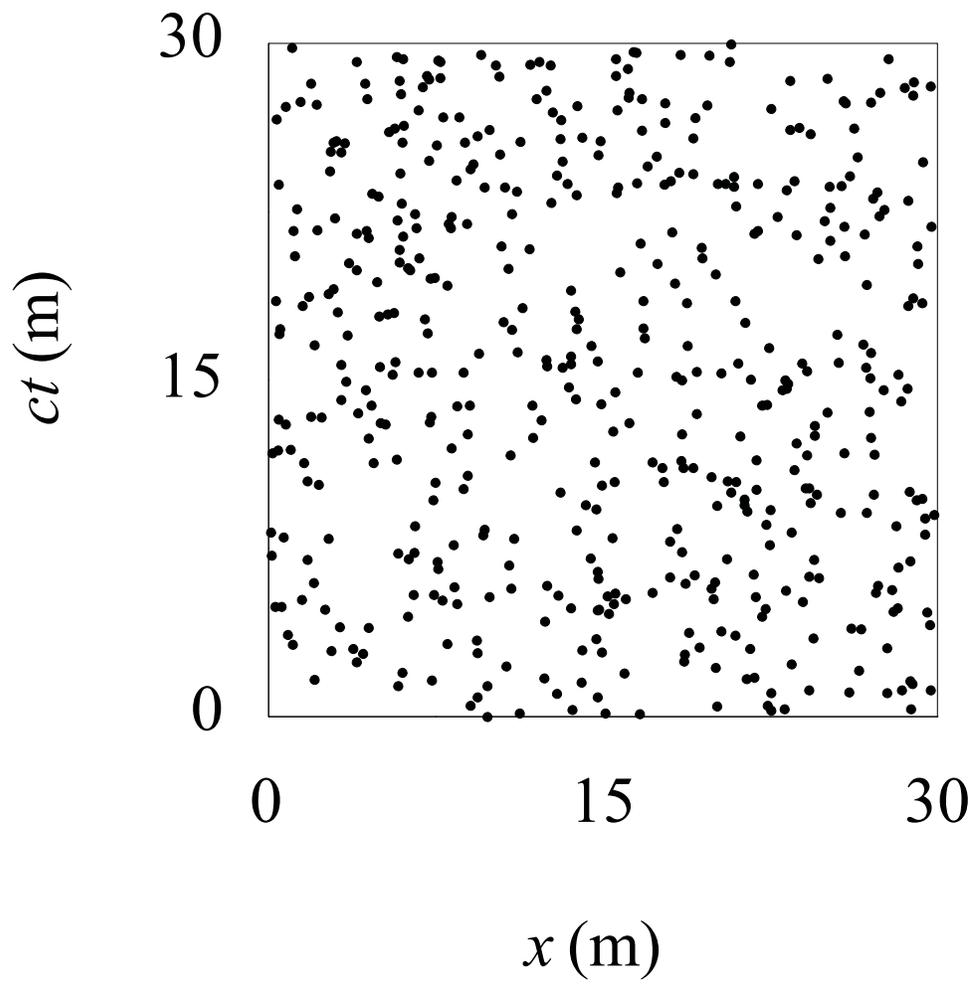
{\bf Figure 1}. A causal set formed from a unit density sprinkling of points
in two-dimensional Minkowski space.

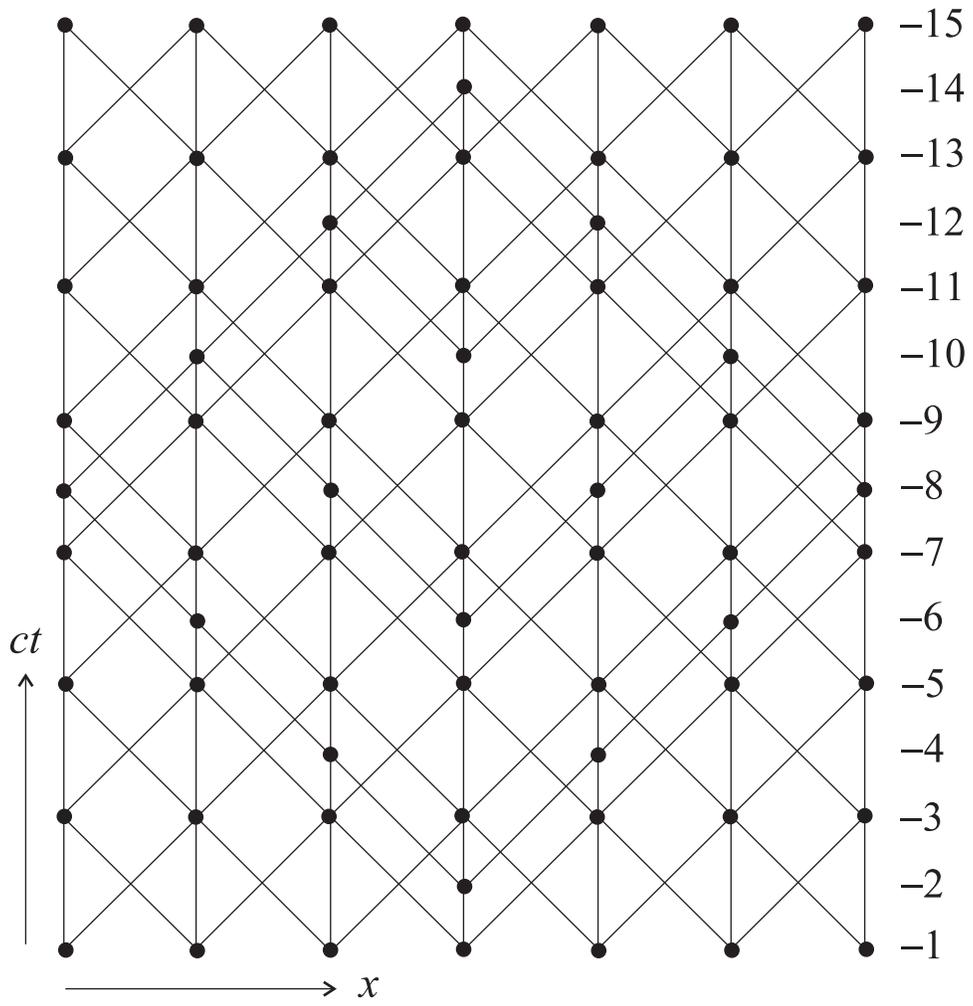
{\bf Figure 2}. A spacetime-Hasse diagram of a two-dimensional causal set.
The dots represent the 72 events in this set and the lines are causal links
between events. This causal set has 15 time steps as enumerated along the
right-hand-side of the figure.

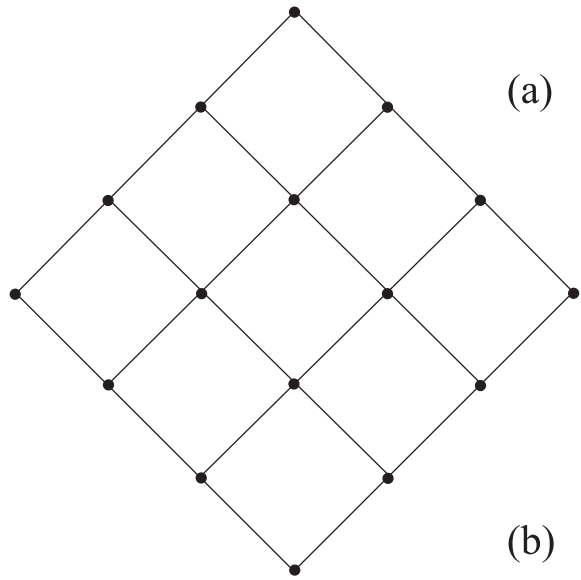
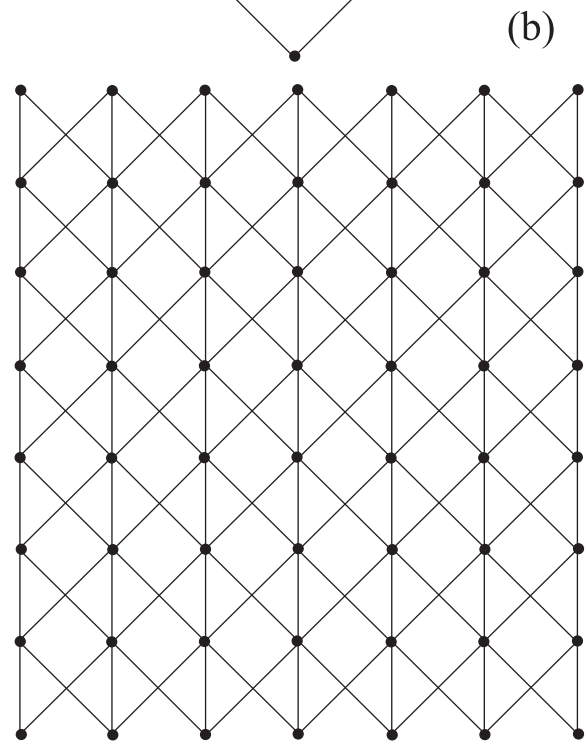
{\bf Figure 3}. Coarse-grainings of the causal set in Fig. 2 formed by
doubling the time scale. (a) The subset formed by the odd time steps only.
(b) The subset formed by the even time steps only. Both coarse-grainings are
more clearly embeddable in two-dimensional Minkowski space than the full
poset in Fig. 2.

\end{document}